\documentclass[a4paper, amsfonts, amssymb, amsmath, reprint, showkeys, nofootinbib, twoside]{revtex4-1}

\usepackage[UKenglish]{babel}
\usepackage[a4paper,centering,hmargin=1.75cm,vmargin=2cm]{geometry} 
\usepackage[utf8]{inputenc}
\usepackage{ulem}
\usepackage{graphicx}
\usepackage{setspace}
\usepackage{lipsum}
\usepackage{physics}
\usepackage{amssymb}
\usepackage{mhchem}
\usepackage{float}
\usepackage{siunitx}
\usepackage{dcolumn}
\usepackage{hyperref}

\DeclareSIUnit\gauss{G}
\newcommand{\yb}{\ce{^{171}Yb^+}~} 
\newcommand{\minus}{\scalebox{0.75}[1.0]{$-$}} 

\begin{document}



\title{Ultrasensitive single-ion electrometry in a magnetic field gradient}

\author{F. Bonus$^{1,2,3\dagger}$, C. Knapp$^{1,\dagger}$, C. H. Valahu$^{1, *}$, M. Mironiuc$^{1,2,3}$, S. Weidt$^{1,3}$ and W. K. Hensinger$^{1,3,\ddagger}$}

\address{\vspace{1mm} $^{1}$Sussex Centre for Quantum Technologies, University of Sussex, Brighton, BN1 9RH, UK}

\address{$^{2}$Department of Physics and Astronomy, University College London, London, WC1E 6BT, UK}

\address{$^{3}$Universal Quantum Ltd, Brighton, BN1 6SB, UK}

\def\thefootnote{$\dagger$}\footnotetext{These authors contributed equally to this work.}
\def\thefootnote{*}\footnotetext{Current address: School of Physics, University of Sydney, NSW 2006, Australia}
\def\thefootnote{$\ddagger$}\footnotetext{Corresponding author: w.k.hensinger@sussex.ac.uk}

\date{\today} 


\begin{abstract}
Hyperfine energy levels in trapped ions offer long-lived spin states. In addition, the motion of these charged particles couples strongly to external electric field perturbations.  These characteristics make trapped ions attractive platforms for the quantum sensing of electric fields. However, the spin states do not exhibit a strong intrinsic coupling to electric fields. This limits the achievable sensitivities. Here, we amplify the coupling between electric field perturbations and the spin states by using a static magnetic field gradient. Displacements of the trapped ion resulting from the forces experienced by an applied external electric field perturbation are thereby mapped to an instantaneous change in the energy level splitting of the internal spin states. This gradient mediated coupling of the electric field to the spin enables the use of a range of well-established magnetometry protocols for electrometry. Using our quantum sensor, we demonstrate AC sensitivities of $\mathrm{S^{AC}_{min}}=\SI{960(10)e-6}{\volt \meter^{-1}\hertz^{-\frac{1}{2}}}$ at a signal frequency of $\omega_{\mathrm{\epsilon}}/2\pi=\SI{5.82}{\hertz}$, and DC sensitivities of $\mathrm{S^{DC}_{min}}=\SI{1.97(3)e-3}{\volt \meter^{-1}\hertz^{-\frac{1}{2}}}$ with a Hahn-echo type sensing sequence. We also employ a rotating frame relaxometry technique, with which our quantum sensor can be utilised as an electric field noise spectrum analyser. We measure electric field signals down to a noise floor of  $\mathrm{S_{E}(\omega)}=\SI{6.2(5)e-12}{\volt^2 \meter^{-2}\hertz^{-1}}$ at a frequency of $\SI{30.0(3)}{k\hertz}$. We therefore demonstrate unprecedented electric field sensitivities for the measurement of both DC signals and AC signals across a frequency range of sub-Hz to $\sim\SI{500}{\kilo \hertz}$. Finally, we describe a set of hardware modifications that are capable of achieving a further improvement in sensitivity by up to six orders of magnitude.
\end{abstract}

\maketitle

Precision measurements of electric fields and forces can be used in a wide range of emergent applications in biological, biomedical, and chemical research~\cite{hanlon2020diamond,aslam2023quantum,krass2022force,yu2021molecular}, particle physics~\cite{yu2021molecular,carney2021mechanical,antypas2022new}, gravitational wave detection~\cite{martynov2016sensitivity}, energy applications~\cite{crawford2021quantum}, and communications~\cite{meyer2018digital,fancher2021rydberg}.
Consequently, a variety of electrometers based on various quantum hardware platforms have been developed, including bulk~\cite{michl2019robust} and single~\cite{dolde2011electric} NV centres, quantum dots~\cite{cadeddu2017electric}, Rydberg atoms~\cite{sedlacek2012microwave,facon2016sensitive,kumar2017atom,jing2020atomic}, and trapped ions in Penning and Paul traps~\cite{gilmore2021quantum,affolter2020phase,shaniv2017quantum,biercuk2010ultrasensitive,blums2018single}.
\par
Existing quantum electrometers have demonstrated ultrasensitive electric field measurements, however they are restricted to certain frequency bands, with few sensors being able to measure sub-kHz frequencies~\cite{jau2020vapor}. This is because commonly used electrometers rely on either near-resonant measurements of transitions within the quantum system~\cite{biercuk2010ultrasensitive,affolter2020phase,facon2016sensitive,jing2020atomic,liu2021phonon}, or resonant pulse techniques on spin states using phase-coherent sensing protocols~\cite{dolde2011electric,michl2019robust,shaniv2017quantum}. In the former, the measurement bandwidth is defined by the frequency of available transitions. In the latter, the lower cut-off frequency of the sensor is constrained by both the achievable coherence times and the coupling strength of the quantum states to the electric-field perturbation, while the upper limit is restricted by the pulse duration of coherent operations on the spin states.
\par 
Access to the frequency band ranging from sub-Hz to several kHz could enable quantum electrometers to be used for a variety of additional applications, including medical imaging techniques such as electrical impedance tomography~\cite{putensen2019electrical}, microscopy~\cite{qiu2022nanoscale}, meteorological applications such as the long-range geolocation of lightning~\cite{said2017towards}, as well as the study of atmospheric phenomena and space weather~\cite{fullekrug2009wideband,silber2017use,hayakawa2021lithosphere}. Geological prospecting techniques are another use case for a low frequency sensor, where applications include the detection of a range of subterranean and submarine features~\cite{reynolds2011introduction,steuer2020comparison}.
\par
In this work, we report on a novel quantum electric field sensor in which a magnetic field gradient is used to couple electric field signals to the energy level separation between the spin states of a two-level system in a single trapped ion. We experimentally demonstrate DC and low frequency AC electric field sensitivities that are unmatched by current state-of-the-art electrometers within our measurement bandwidth. We also demonstrate the versatility of our sensing scheme by employing a magnetometry technique to measure electric field noise.\par 

\begin{figure*}[]
    \centering
    \includegraphics[width=.9\textwidth]{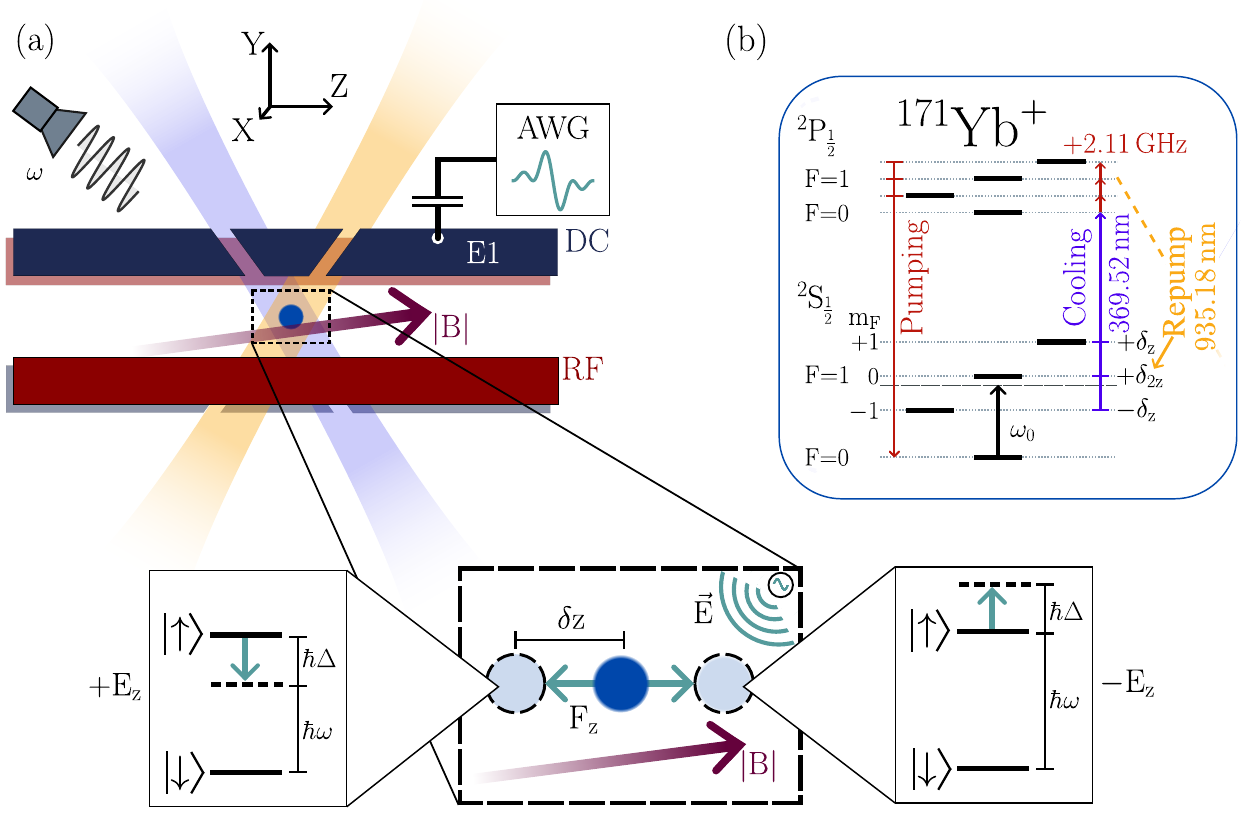}
    \caption{\textbf{Electric field sensing with a trapped ion in a magnetic field gradient} 
    \textbf{a.} A single ion is confined in an RF Paul trap. Segmented DC electrodes (blue) provide confinement in the axial (z) direction, whilst the RF electrodes (red) provide confinement in the radial (x,y) directions. A magnetic field gradient of $\frac{\partial \mathrm{B}}{\partial \mathrm{z}}=\SI{22.41(1)}{\tesla \per \meter}$ is present along z. Doppler cooling and re-pump lasers at wavelengths of \SI{369.52}{\nano \meter} and \SI{935.18}{\nano \meter} respectively are indicated by the blue and orange beams. Transitions between the internal spin states are driven using an external microwave emitter. Electric field signals are applied to the ion through a DC end-cap electrode (E1), and are generated using an AWG that is capacitively coupled onto the external signal chain of E1. \textbf{Zoom:} An external electric field $\vec{\mathrm{E}}$ applies a force $\vec{\mathrm{F}}$ on the ion, resulting in a displacement $\delta \mathrm{z}$. The transition frequency of the spin states is then shifted by $\Delta$ due to the magnetic field gradient.
    \textbf{b.} Simplified energy level diagram of the \yb{} ion. Doppler cooling, optical pumping and state detection are carried out using the standard resonance fluorescence scheme described in \cite{olmschenk2007manipulation}. Phase coherent operations on the second order magnetic field sensitive $\ket{\mathrm{F=0,m_F=0}}$ to $\ket{\mathrm{F=1,m_F=0}}$ transition and first order sensitive $\ket{\mathrm{F=0,m_F=0}}$ to $\ket{\mathrm{F=1,m_F=+1}}$ transition are driven by resonant microwave fields.}
    \label{fig-sensing_scheme}
\end{figure*}

\begin{figure*}[]
    \centering
    \includegraphics[width=0.84\textwidth]{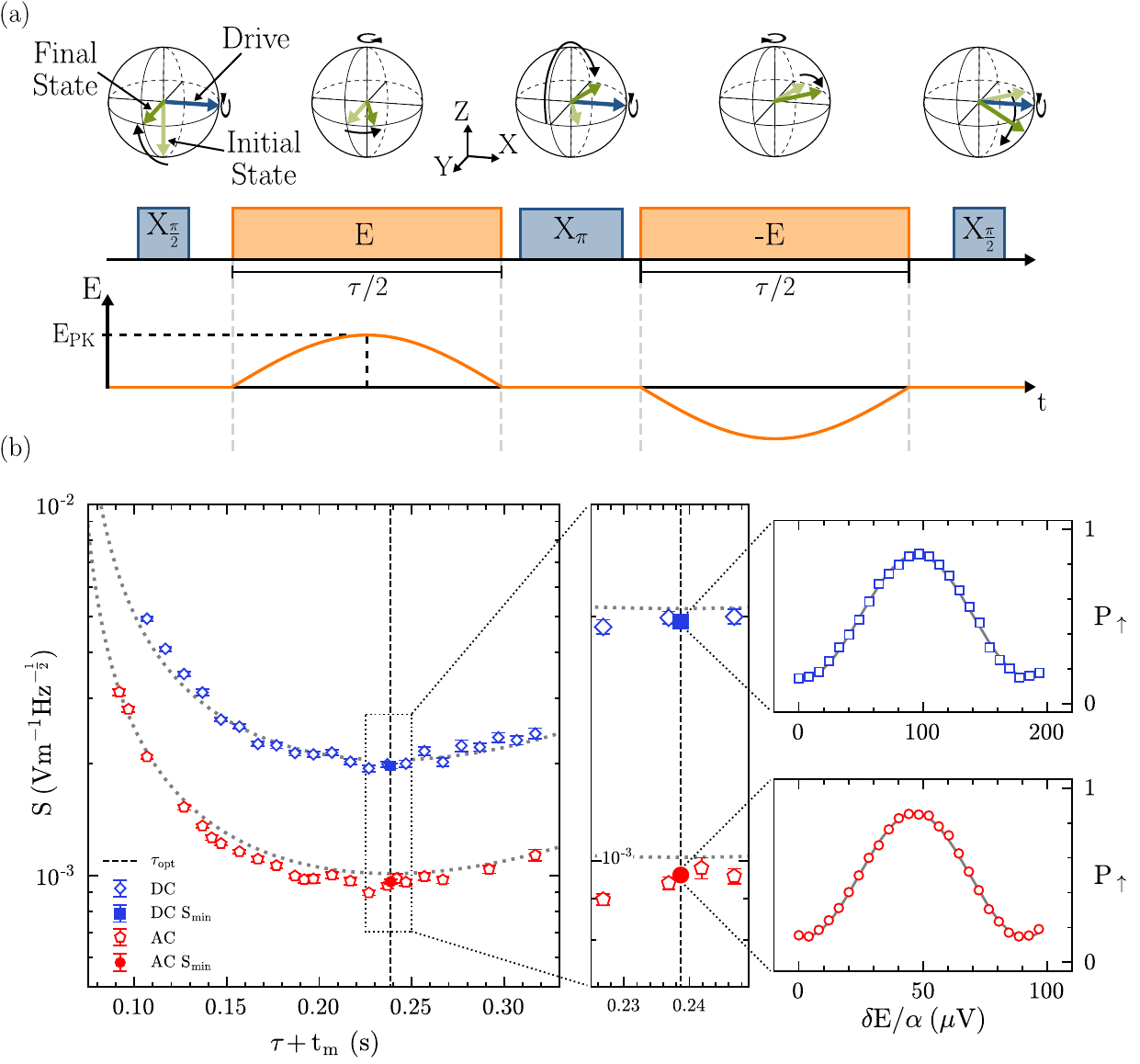}
    \caption{\textbf{Measuring AC and DC sensitivities.} \textbf{(a)} Bloch sphere representation of the quantum state evolution, pulse sequence diagram and plot of the evolution of the electric field amplitude at the ion $\mathrm{E}$ for the AC sensing technique. Blue arrows and rectangles represent the microwave drive, and the orange rectangles and lines represent interaction with the electric field. The initial and final spin states are shown in light and dark green respectively. Each electric field interaction period of duration $\tau/2$ features a half oscillation of a signal with frequency $\omega_{\epsilon}/2\pi=1/\tau$. \textbf{(b)} Sensitivity of AC and DC sensing sequences against shot duration, $\tau + \mathrm{t_m}$ for evolution times  ranging from $\tau=\SI{25}{\milli \second}$ to $\tau=\SI{250}{\milli \second}$, corresponding to signal frequencies of $\omega_{\epsilon}/2\pi=\SI{40}{\hertz}$ to $\omega_{\mathrm{\epsilon}}/2\pi=\SI{4}{\hertz}$. Measurements of the sensitivity near the optimal evolution time $\tau_{\mathrm{opt}}=\SI{172(2)}{\milli \second}$ are indicated by the square and circular markers in the central insets for DC and AC respectively, and are sampled with 2950 (DC) and 3750 (AC) shots. The measured probability $\mathrm{P_{\uparrow}}$ against applied electrode voltage, $\delta \mathrm{E}/\alpha$ where $\delta \mathrm{E}=\frac{2}{\pi}\mathrm{E_{PK}}$, for DC (AC) sensitivity measurements at $\mathrm{S_{min}}$, along with a least squares fit to a sine wave (solid grey), are shown in the upper (lower) right insets. The dotted grey lines on the main plot are the theoretically expected curves for DC and AC sensing from equation~\ref{eq-min-sensitivity-theory}.}
    \label{fig-sensitivity}
\end{figure*}

\section*{Results}
\label{sec:results}
We consider a single ion with charge $\mathrm{q}$ confined in a radio-frequency (RF) Paul trap. A magnetic field gradient is applied at the position of the ion, as depicted in figure~\ref{fig-sensing_scheme}. An electric field perturbation, $\delta \vec{\mathrm{E}}(\mathrm{t})$, will alter the confining potential and exert a force $\delta\vec{\mathrm{F}}\mathrm{(t)}=\mathrm{q}\delta\vec{\mathrm{E}}\mathrm{(t)}$ on the ion. This force displaces the ion along the vector $\vec{\mathrm{r}} = \mathrm{(r_x, r_y, r_z)}$ by an amount (see methods)
\begin{equation}
\label{eq:transduction_parameter}
\delta \mathrm{r_i(t)} = \frac{\mathrm{q}}{\mathrm{m} \nu^2_\mathrm{i}} \delta \mathrm{E_{i}(t)},
\end{equation}
where $\mathrm{i \in \{x, y, z\}}$, and $\mathrm{m}$ and $\mathrm{\nu_i}$ are the mass of the ion and its vibrational frequency along the i-axis respectively. 
The displacement $\delta\mathrm{r_i}$ of the trapped ion causes a change $\Delta$ in the transition frequency $\omega$ of its spin states due to the position dependent Zeeman shift. The transduction parameter, $\gamma_\mathrm{i}$, defines the sensitivity of the spin state transition frequency to changes in the electric field and is given by
\begin{equation}
\label{eq-gamma}
\gamma_\mathrm{i} = \frac{\partial \omega}{\partial \mathrm{E_i}} = \frac{\partial \omega}{\partial \mathrm{B}} \frac{\partial \mathrm{B}}{\partial \mathrm{r_{i}}} \frac{\partial \mathrm{r_{i}}}{\partial \mathrm{E}_\mathrm{i}},
\end{equation}
where $\frac{\partial \omega}{\partial \mathrm{B}}$ is the sensitivity of the transition frequency to changes in magnetic field, $\frac{\partial \mathrm{B}}{\partial \mathrm{r_{i}}}$ is the strength of the magnetic field gradient along $\mathrm{r}_\mathrm{i}$, and $\frac{\partial \mathrm{r_i}}{\partial \mathrm{E}_\mathrm{i}} = \mathrm{q}/\mathrm{m} \nu^2_\mathrm{i}$ is the change in position for a given change in electric field at the ion. Equation~\ref{eq-gamma} highlights the mechanism of our sensing scheme; the magnetic field gradient transforms electric fields into magnetic fields in the reference frame of the ion, which allows for the implementation of a wide range of magnetometry techniques for electrometry. From equation~\ref{eq-gamma}, we see that a stronger coupling is achieved by lowering the vibrational frequency of the ion, increasing the strength of the magnetic field gradient, using ions with a larger charge-to-mass ratio, or by employing transitions with a higher sensitivity to magnetic fields. Electric field vector sensing is also in principle possible by tuning the confinement strength of the ion trap to maximise $\gamma_\mathrm{i}$ along one axis, whilst suppressing it along the others.\par
All experimental demonstrations of our sensing scheme were conducted using a single \yb{} ion confined in a linear RF blade-trap with segmented DC electrodes~\cite{lake_2015}. A magnetic field gradient of $\frac{\partial \mathrm{B}}{\partial\mathrm{z}}=\SI{22.41(1)}{\tesla \per \meter}$ is generated along the axial (z) direction of the trap by a set of samarium-cobalt magnets. The magnetic field strength at the unperturbed ion position is $\mathrm{B}_0=\SI{8.3767(4)}{\gauss}$. Doppler cooling and re-pump lasers, with wavelengths of \SI{369.52}{\nano \meter} and \SI{935.18}{\nano \meter} respectively, are used to cool the ion to near the Doppler limit, whilst coherent operations on the spin states are realised by applying microwave fields using an external microwave emitter as shown in figure~\ref{fig-sensing_scheme}. Further details of the experimental setup and control techniques can be found in the methods section. Electric field signals are generated using an AWG, and injected onto one of the DC end-cap electrodes of the ion trap by capacitively coupling across a $\SI{220}{pF}$ capacitor (see methods). The applied electric field strength is characterised by a geometric factor, $\mathrm{\alpha_i = \frac{\partial E_i}{\partial V}}$, which relates the electric field at the position of the ion to the DC voltage applied to the electrode. Strong radial confinement ($\mathrm{\nu_x \approxeq \nu_y} \approx \SI{1.5}{\mega \hertz}$) suppresses coupling of radial electric field components to the spin state transition frequency. The subsequent experiments therefore measure solely the axial (z) component of the electric field, where we find $\alpha_\mathrm{z}=\alpha = -95.64(4)$ (see methods), and we will drop the subscript from here on.\par 
\subsection*{AC and DC Sensing}
\label{sec-AC-sensing}
We use the $\ket{\downarrow}=\ket{\mathrm{F}=0,\mathrm{m_F}=0}$ and $\ket{\uparrow}=\ket{\mathrm{F}=1,\mathrm{m_F}=0}$ energy levels of the $^2\mathrm{S}_{\frac{1}{2}}$ hyperfine manifold of \yb{} for the measurements of AC and DC fields (see figure~\ref{fig-sensing_scheme}(b)). The energy level separation of the spin states is a function of the magnetic field at the ion, and is given by $\omega=\omega_0 + \delta_{2\mathrm{z}}$ where $\omega_0/2\pi \approx \SI{12.64}{\giga \hertz}$ is the hyperfine splitting at zero magnetic field and $\delta_{2\mathrm{z}}/2\pi = \SI[parse-numbers=false]{310.8 \mathrm{B}^2}{\hertz}$ (B in Gauss) is the second order Zeeman splitting~\cite{Vanier1989}. The vibrational frequency along z is measured to be $\nu_z/2\pi=\SI{161.191(8)}{\kilo \hertz}$, from which we calculate the transduction parameter $\gamma = \SI{3998(2)}{\radian \meter \per \volt}$ (see methods).\par
The sensitivity to AC signals is characterised using a Hahn-echo type sequence, where the electric field signal is applied during the free procession time $\tau$ as described in~\cite{degen_2017} and illustrated in figure~\ref{fig-sensitivity}(a). We apply an AC electric field with a frequency $\omega_{\epsilon}=\tau^{-1}$. The pulse sequence maps the electric field amplitude onto the probability of finding the spin in the $\ket{\uparrow}$ state, $\mathrm{P}_\uparrow$. The displacement of the ion in the magnetic field gradient results in an instantaneous field-induced detuning, $\Delta$, of the two-level system transition frequency. A superposition of the spin states will therefore experience a phase shift of $\mathrm{d}\phi = \Delta\mathrm{(t)dt}$, where $\Delta\mathrm{(t)}=\gamma \delta\mathrm{E(t)}$ is the detuning of the spin transition frequency. The total accumulated phase over the signal duration $\tau$ is $\phi=\int_0^\frac{\tau}{2}  \Delta \mathrm{(t) dt}-\int_\frac{\tau}{2}^\tau \Delta \mathrm{(t) dt}$, which is a function of the electric field amplitude $\delta\mathrm{E(t)}$ and $\tau$. The electric field amplitude is linearly increased for each interaction time $\tau$, leading to sinusoidal oscillations in the $\mathrm{P}_\uparrow$. A linear least squares fit is then used to fit an equation of the form $\mathrm{P_\uparrow}=\frac{1}{2}+\frac{\mathrm{A}}{2}\sin{(\frac{2\pi}{\mathrm{\kappa}}\mathrm{E})}$ to the data. Here $\mathrm{A}$ is the fringe amplitude, $\mathrm{\kappa}$ is the electric field required to induce a $2\pi$ phase rotation of the spin, and $\mathrm{E}$ is the electric field at the ion. We extract the resulting maximal derivative $\frac{\partial \mathrm{P}_{\uparrow}}{\partial \mathrm{E}}$ and use this to calculate the minimum detectable electric field 
\begin{equation}
\label{eq-emin}
\mathrm{E_{min}}=\sigma_{\mathrm{tot}} \left(\frac{\partial \mathrm{P_{\uparrow}}}{\partial \mathrm{E}}\right)^{-1},
\end{equation}
%
%
where $\mathrm{\sigma_{tot}}$ is the total readout uncertainty due to quantum projection noise ($\sigma^2_{\mathrm{quantum}}$) and classical readout noise ($\sigma^2_{\mathrm{readout}}$), given by $\sigma^2_{\mathrm{tot}}=\sigma^2_{\mathrm{quantum}}+\sigma^2_{\mathrm{readout}} \approxeq 1/(\mathrm{4C^2N})$~\cite{degen_2017}. Here, $\mathrm{C}\approx1/\sqrt{(1+4\eta)}$ is an overall readout efficiency parameter~\cite{taylor2008high}, $\mathrm{N}$ is the number of measurements of the spin state, and $\eta$ is the infidelity associated with state preparation and measurement (SPAM). We measure a SPAM infidelity of $\eta=1.8\times10^{-2}$, resulting in $\mathrm{C} = 0.97$. The sensitivity, defined as the minimum detectable signal measured over one second of averaging, is then calculated as $\mathrm{S}=\mathrm{E_{min}}\sqrt{\mathrm{t_{exp}}}$. Here, $\mathrm{t_{exp}} = \mathrm{N}(\tau + \mathrm{t_m})$ is the total experimental duration, where $\mathrm{t_m}$ is the overhead associated with initialisation, manipulation, and readout of the sensor. From~\cite{degen_2017}, the optimum sensitivity for a given evolution time $\tau$ is
\begin{equation}
\label{eq-min-sensitivity-theory}
\mathrm{S_{min}}=\frac{\mathrm{e}^{\chi(\tau)}\sqrt{\tau+\mathrm{t_{m}}}}{\gamma \mathrm{C} \tau},
\end{equation}
where $\chi(\tau)$ is the associated decoherence function of the two-level system.  The measured sensitivity for each evolution time is shown in figure~\ref{fig-sensitivity}(b). AC waveforms are applied across the capacitor for various evolution times. These waveforms are pre-compensated to account for the frequency dependent phase offset induced by the capacitor (see methods). \par
Whilst DC signals cannot be injected across the input capacitor, the sensitivity of the sensor to DC electric fields can still be characterised by injecting a time varying signal. We also employ a Hahn-echo type sequence for DC sensing, where the interaction between the electric field and the sensor only occurs during the first half ($\tau/2$) of the total free evolution time (see methods). The average electric field over the course of this half oscillation is given by $\mathrm{\bar{E} = \frac{2}{\pi} E_{PK}}$, where $\mathrm{E_{PK}}$ is the electric field amplitude. Correspondingly, the sensor accumulates the same amount of coherent phase $\phi$ as if it were evolved under a square DC pulse of amplitude $\mathrm{E_{DC}=\bar{E}}$.\par
The data shown in figure \ref{fig-sensitivity} are in good agreement with the theory, which is plotted from equation~\ref{eq-min-sensitivity-theory}. For AC sensing, we see a $\sim 5\%$ offset of the sensitivity relative to the theory near $\tau_\mathrm{opt}$. This is due to higher frequency electric field components capacitively coupling onto the electrode, which could not be fully eliminated by the pre-compensation sequence.\par
The local minimum in sensitivity $\mathrm{S_{min}}$ occurs at an optimal evolution time $\tau_\mathrm{opt}$. This is because the electric field induced phase accumulation increases linearly with $\tau$, but is counteracted by the reduction in fringe-contrast of the quantum system due to decoherence which follows a Gaussian functional form. $\tau_\mathrm{opt}$ can therefore be determined from equation \ref{eq-min-sensitivity-theory}. Experimentally, we find the local minimum in sensitivity to be at $\tau_{\mathrm{opt}}=\SI{172(2)}{\milli \second}$ for $\mathrm{t_m}=\SI{66.839}{\milli \second}$, and the coherence time $\mathrm{T}_{2}=\SI{304(3)}{\milli \second}$  (see methods). We measure a minimum AC sensitivity of $\mathrm{S^{\mathrm{AC}}_{min}}=\SI{960(10)e-6}{\volt \meter^{-1}\hertz^{-\frac{1}{2}}}$ at a signal frequency of $\omega_\mathrm{\epsilon}=\mathrm{ \tau^{-1}_{opt} = \SI{5.82}{\hertz}}$, and a minimum DC sensitivity of $\mathrm{S^{\mathrm{DC}}_{min}}=\SI{1.97(3)e-3}{\volt \meter^{-1} \hertz^{-\frac{1}{2}}}$.\par 
\begin{figure}
    \centering
    \includegraphics[width=\columnwidth]{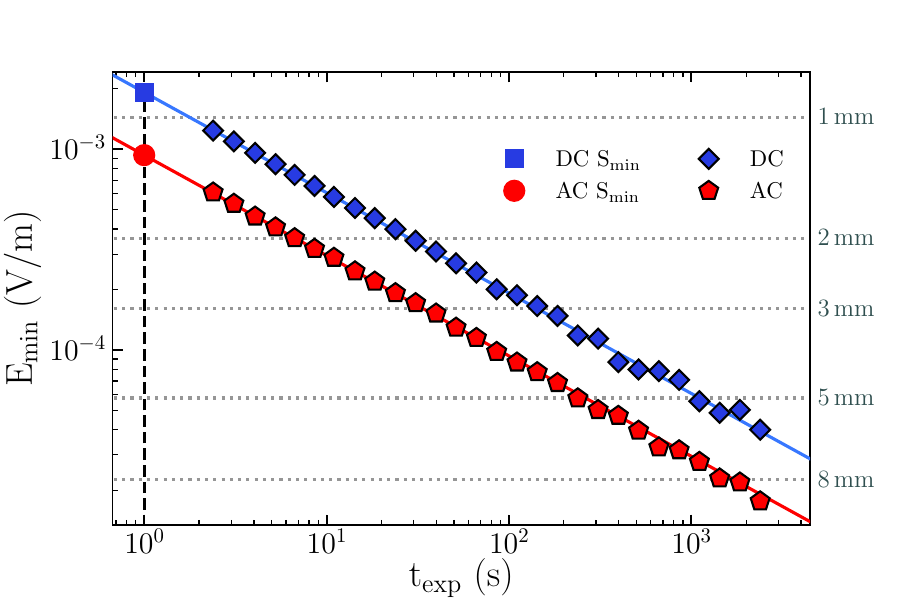}
    \caption{\textbf{Minimum detectable signal against measurement time.}
    Measured values of $\mathrm{E_{min}}$ at fixed measurement times for DC (AC) sensing are shown in blue (red). The blue (red) lines show the theoretical dependence of $\mathrm{E_{min}}$, limited only by quantum projection noise. The value of $\mathrm{E_{min}}$ for a measurement time of $\SI{1}{\second}$ (which defines the minimum sensitivity of the quantum sensor) is also shown (dashed black). The classical readout error is approximately equivalent for measurements on the $\ket{\downarrow}$ and $\ket{\uparrow}$ states, meaning it does not contribute to the experimentally measured standard deviation shown in this figure. The dotted grey lines represent the magnitude of electric field emanating from a single elementary charge at the indicated distance.
    }
    \label{fig-emin-measurement}
\end{figure}
\begin{figure*}[]
    \centering
    \includegraphics[width=0.9\textwidth]{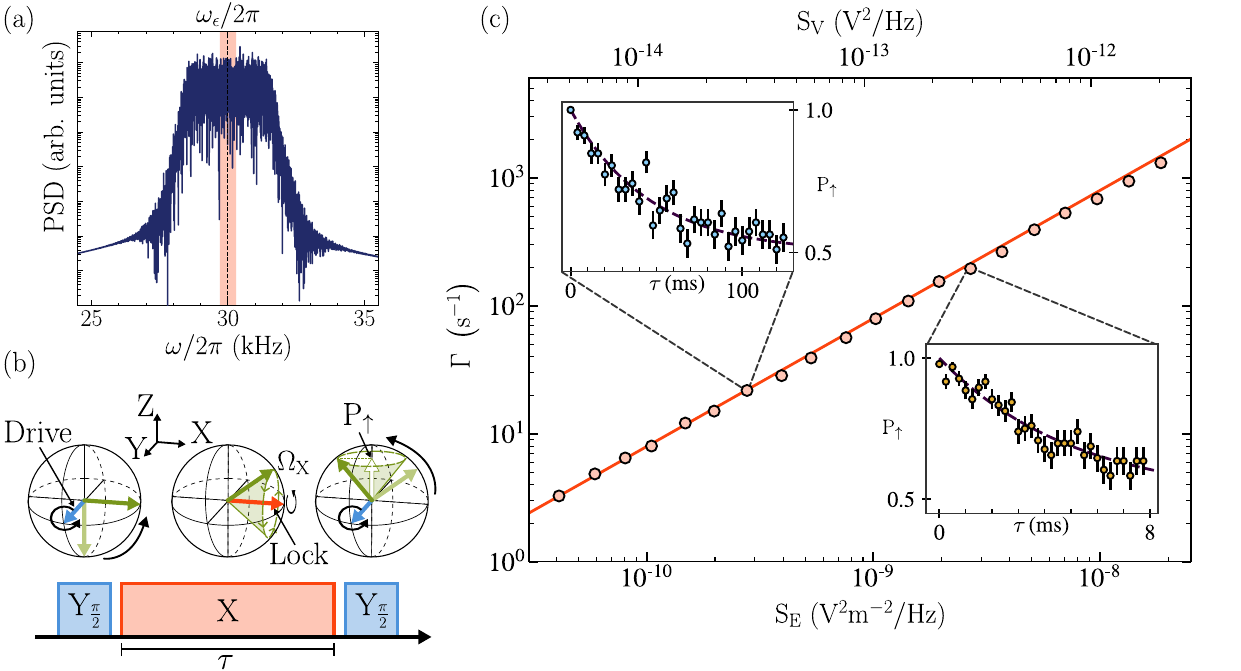}
    \caption{\textbf{Rotating frame relaxometry for electric field sensing} \textbf{(a)} Periodogram of the applied noise. White noise with a bandwidth $\mathrm{B}/2\pi= \SI{3}{\kilo \hertz}$ centred around $\omega/2\pi = \SI{30}{\kilo \hertz}$ is applied into the system for various signal PSDs. The Rabi frequency $\omega_\mathrm{\epsilon}=\SI{30.0}{\kilo \hertz}$ of the spin-locking pulse is indicated by the black dashed line, and the shaded region indicates the $1\sigma= \pm\SI{300}{\hertz}$ error of the Rabi frequency measurement. \textbf{(b)} Pulse sequence diagram and associated Bloch-sphere representation of the spin-locking sequence. A $\mathrm{Y}_{\frac{\pi}{2}}$-pulse aligns the spin state with the X-axis. An X-pulse with Rabi frequency $\Omega_\mathrm{X}$ locks the state vector to the X-axis. Resonant noise at the spin-locking Rabi frequency drives the $\ket{+}\rightarrow\ket{-}$ transition incoherently. A final $\mathrm{Y}_{\frac{\pi}{2}}$-pulse transfers state population into the $\sigma_z$-basis for readout. The outer radius of the cone represents all possible alignments of the final state vector. The measured probability over many shots $\mathrm{P}_\uparrow$ is represented by the projection of the vector onto the $\mathrm{z}$-axis (white vector). \textbf{(c)} Measurement of decay rate $\Gamma$ against the resonant voltage power spectral density of applied noise. Round markers indicate fits of probability measurements to exponential decay functions, error bars are within the size of the marker. The solid line is given by equation \ref{eq_gamma_sv}. The left (right) inset shows measurements of the decay rate and the associated fit for a PSD of $\mathrm{S_E}=\SI{2.770e-10}{\volt^2 \meter^{-2}\hertz^{-1}}$ ($\mathrm{S_E}=\SI{2.689e-9}{\volt^2 \meter^{-2}\hertz^{-1}}$) resulting in a decay rate of $\Gamma=\SI{22(1)}{\per \second}$ ($\Gamma=\SI{195(9)}{\per \second}$).}
    \label{fig-splk}
\end{figure*}
In order to determine if our quantum sensor is shot noise limited, $\mathrm{M}=275000$ measurements (shots) are taken at the optimal evolution time $\tau_{\mathrm{opt}}$ for both AC and DC signals. The electric field amplitude is set so that a measurement of the quantum system yields a probability of $\mathrm{P}_{\uparrow}=0.5$. The set of M shots is then subdivided into $\mathrm{k=M/N}$ sets of $\mathrm{N}$ shots. From this, we calculate $\mathrm{k}$-individual means, corresponding to the mean probability of each set of $\mathrm{N}$ shots. 
Using equation \ref{eq-emin} we plot the minimum electric field $\mathrm{E_{min}}$ calculated using the standard deviation of each set of $\mathrm{k}$ means (equation \ref{eq-emin})  against $\mathrm{N}$ in figure~\ref{fig-emin-measurement}, by varying the total measurement duration, $\mathrm{t_{exp}}$ which is a function of $\mathrm{N}$.
The measurement shows that the minimum detectable electric field follows a $\mathrm{1/\sqrt{t_{exp}}}$ dependence, which is consistent with a shot noise limited sensor.
We find that for one second of integration time of an AC signal, the quantum electrometer is able to measure a minimum detectable electric field equivalent to an elementary charge at a distance of $\SI{1.225(6)}{\milli \meter}$.\par
%
%
\subsection*{Rotating frame relaxometry}
\label{sec-relaxo}
In the previous section, we have shown the measurement of DC signals and AC signals at well-defined frequencies and phases. Our sensor can however also be employed for the measurement of stochastic signals that feature a discontinuous phase evolution over the measurement interval. We demonstrate this by using our sensing scheme to measure power spectral densities of electric field noise. This is done using a spin-locking sequence. This technique is well established in magnetometry~\cite{ithier2005decoherence,yan2013rotating}, however the gradient mediated coupling of our scheme enables the first implementation of spin-locking to measure electric field noise. The pulse sequence, outlined in figure~\ref{fig-splk}(b), begins by initialising the spin state into the $\ket{\mathrm{+X}}$ eigenstate. A resonant drive of the form $\frac{\omega_\epsilon}{2}\sigma_\mathrm{x}$, with Rabi frequency $\omega_\epsilon$, is applied parallel to the orientation of the spin state, locking the spin along the x-axis of the Bloch sphere. The resonant interaction lifts the degeneracy of the $\ket{\pm \mathrm{X}}$ eigenstates by an energy $\epsilon=\hbar \omega_\epsilon$, thereby making the two-level system only sensitive to $\sigma_{\mathrm{Z}}$-type signals oscillating at angular frequency $\epsilon/\hbar = \omega_\epsilon$, effectively creating a quantum electric field noise spectrum analyser. In the presence of electric field noise, the resonant drive is applied for a duration $\tau$, after which the spin state is mapped into the $\sigma_\mathrm{z}$ basis for detection. The measured probability follows an exponential decay in time of the form
\begin{equation}
\label{eq_exp_decay}
\mathrm{P}_{\uparrow}=\frac{1}{2}\left(1+e^{-\tau \Gamma} \right),
\end{equation}
where $\Gamma$ is the decay rate of the system. The measured decay is a result of electric field noise at angular frequency $\omega_\epsilon$ being transformed into $\sigma_\mathrm{z}$-noise on the spin states through the coupling induced by the magnetic-field gradient. We define the power spectral density (PSD) of electric field noise at an arbitrary angular frequency $\omega$ as $\mathrm{S_\mathrm{E}(\omega)=\int_{-\infty}^{+\infty} \langle \delta E(0) \delta E(t)\rangle e^{i\omega t}dt}$. The corresponding PSD of $\sigma_\mathrm{z}$-noise is then related to the PSD of electric field noise by $\mathrm{S_{z}}(\omega_\epsilon)=\gamma^2 \mathrm{S_E}(\omega_\epsilon)$, giving a spin-locking decay rate of~\cite{yan2013rotating}\par
\begin{equation}
\label{eq_gamma_sv}
\Gamma=\frac{1}{2}\mathrm{S_{z}}(\omega_\epsilon).
\end{equation}
Equations \ref{eq_exp_decay} and \ref{eq_gamma_sv} therefore make it possible to extract the PSD of electric field noise at the angular frequency of the resonant drive $\omega_\epsilon$.\par
To characterise our sensor experimentally, we capacitively inject electric field noise into the system for the duration of the spin-locking drive pulse. The waveform comprises white noise in a \SI{3}{\kilo \hertz} bandwidth centred around the resonant drive frequency $\Omega_\mathrm{X}/2\pi = \SI{30.0(3)}{\kilo \hertz}$, as illustrated in figure~\ref{fig-splk}(a).
\par 
For this experiment, we use the first order magnetic field sensitive $\ket{\downarrow}=\ket{\mathrm{F}=0,\mathrm{m_F}=0}$ and $\ket{\uparrow}=\ket{\mathrm{F}=1,\mathrm{m_F}=1}$  spin states, where the transition frequency is $\omega=\omega_0 + \delta_\mathrm{z}$ and $\delta_\mathrm{z}/2\pi=\SI{1.4}{\mega \hertz \per \gauss}$ is the first-order Zeeman shift. In addition, we set the axial secular frequency to $\nu_z/2\pi=\SI{264.79(1)}{\kilo \hertz}$, from which we calculate a coupling strength of $\gamma = \SI{398.6(2)e3}{\radian \meter \volt^{-1}}$. We first verify the relation of equation~\ref{eq_gamma_sv} by characterising the decay rate $\Gamma$ for various injected noise amplitudes (see figure.~\ref{fig-splk}(c)). This is done by measuring $\mathrm{P}_\uparrow$ as a function of spin-locking drive durations $\tau$, and fitting the data to equation~\ref{eq_exp_decay}. We then characterise the minimum detectable signal, which is defined as the electric field PSD for which the signal-to-noise ratio (SNR) is equal to 1. The SNR is calculated by measuring the decay rate in the absence of injected noise. From this, we measure a decay rate $\Gamma_0=\SI{0.49(4)}{\per \second}$, corresponding to a minimum detectable signal of $\mathrm{S^{min}_E}=\SI{6.2(5)e-12}{\volt^2 \meter^{-2} \hertz^{-1}}$. \par 
%
%
\section*{Discussion}\label{sec-discuss}
\begin{figure}
    \centering
    \includegraphics[width=\columnwidth]{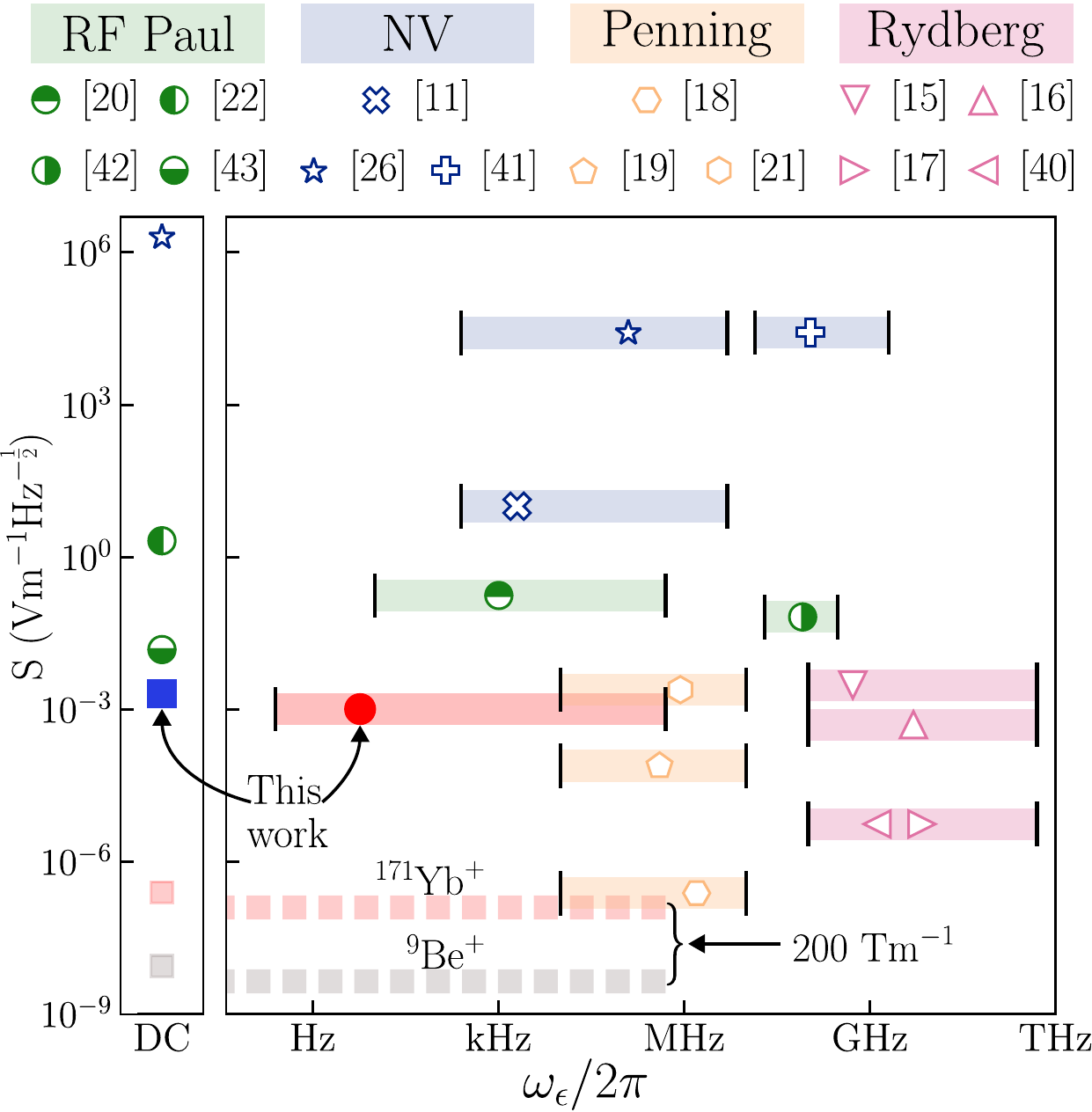}
    \caption{\textbf{Sensitivity and bandwidth comparison of quantum electrometers.} Comparison of electric field sensitivities and bandwidths of various quantum sensing hardware platforms. Markers show measured sensitivities as described in the corresponding reference, and shaded regions illustrate the approximate bandwidths. The blue square and red circle markers represent the measured sensitivities described in this work. The lower cutoff frequency of the experimental system is estimated for a coherence time limited by magnetic field noise at large $\nu$ corresponding to a cutoff frequency of \SI{0.25}{\hertz}. The dashed lines indicate the estimated achievable sensitivities for a system with $\nu/2\pi=\SI{100}{\kilo \hertz}$ and $\frac{\partial \mathrm{B}}{\partial \mathrm{z}}=\SI{200}{\tesla \meter^{-1}}$ using a first order magnetic-field sensitive state of \yb{} (light red) and \ce{^{9}Be^+} (light grey). Penning \cite{biercuk2010ultrasensitive,affolter2020phase,gilmore2021quantum}, Rydberg \cite{jing2020atomic,facon2016sensitive,holloway2022rydberg,kumar2017atom}, NV \cite{michl2019robust,qiu2022nanoscale,cheng2023radio}, Paul \cite{shaniv2017quantum,blums2018single,wu2023quantum,wei2022detection}}
    \label{fig-sens-comp}
\end{figure}
We report on a novel quantum sensing technique for trapped ions in RF traps. A magnetic field gradient is used to couple electric field induced displacements of the ion to its spin state energy level splitting, enabling the use of magnetometry protocols for electrometry. We demonstrate our scheme with a single trapped \yb{} ion by measuring the axial component of electric field signals emitted by an in-vacuum electrode. We measure AC sensitivities of $\mathrm{S^{AC}_{min}}=\SI{960(10)e-6}{\volt \meter^{-1}\hertz^{-\frac{1}{2}}}$ for a signal frequency of $\tau^{-1}=\SI{5.82}{\hertz}$, and DC sensitivities of $\mathrm{S^{DC}_{min}}=\SI{1.97(3)e-3}{\volt \meter^{-1}\hertz^{-\frac{1}{2}}}$. In addition, we employ a spin-locking sequence to measure stochastic signals which feature a discontinuous phase evolution over the measurement time. We determine a minimum detectable electric field PSD of $\mathrm{S_{E}(\omega)}=\SI{6.2(5)e-12}{\volt^2 \meter^{-2}\hertz^{-1}}$ at a frequency of $\omega/2\pi=\SI{30.0(3)}{\kilo \hertz}$.\par
In figure~\ref{fig-sens-comp} we compare the sensitivity and bandwidth of our scheme with current state-of-the-art quantum electrometers. Current quantum hardware platforms use a variety of measurement schemes for electrometry, resulting in a range of achievable bandwidths and measurable sensitivities. Single~\cite{qiu2022nanoscale,cheng2023radio} and bulk NV centres~\cite{michl2019robust} utilise resonant pulse schemes on their spin transition frequency and are able to operate at ambient conditions, allowing highly increased flexibility in sensor placement~\cite{krass2022force}. However, coherence times and coupling strengths limit both the achievable sensitivities and the bandwidth. Rydberg atoms measure Stark shifts on internal transitions induced by near-resonant fields, enabling high sensitivity electrometry in the \SI{100}{\mega \hertz} to \SI{500}{\giga \hertz} range~\cite{facon2016sensitive,sedlacek2013atom}. Ion crystals in Penning traps are sensitive to electric fields at or near the motional resonances of the crystal, which are typically in the \SI{50}{\kilo \hertz} to \SI{10}{\mega \hertz} range~\cite{biercuk2010ultrasensitive,affolter2020phase,gilmore2021quantum}. Rydberg and Penning trap architectures have also demonstrated electric field sensitivities below the standard quantum limit (SQL) through entanglement-based schemes~\cite{gilmore2021quantum,facon2016sensitive}.  Finally, there exist a variety of sensors based on RF Paul traps, which implement both fluorescence based schemes to measure DC electric fields~\cite{blums2018single,wei2022detection,liu2022single}, and resonant pulse schemes for Doppler shift measurements~\cite{shaniv2017quantum}.\par
The achieved minimum sensitivities discussed in this work are unmatched by existing sensing hardware platforms across the measurement bandwidth of our sensor. Our sensing scheme can be used for highly sensitive electric field measurements in the DC and sub-Hz to $\sim \SI{500}{\kilo \hertz}$ frequency range. The lower cut-off frequency is limited by the coherence time of the two-level system, whilst the upper cut-off frequency is a function of the maximal achievable Rabi frequency of the refocussing $\pi$-pulses. Our experimentally measured optimal sensitivity is limited by both classical noise, and hardware constraints specific to the experimental system. Voltage noise on the electrodes of the ion trap directly couples to the spin states, which limits the $\mathrm{T_2}$ coherence time. Previous measurements on our particular experimental setup have shown that the coherence time of our system is dominated by voltage noise on the trapping electrodes and scales as $\mathrm{T_2} \propto \nu^4\left(\frac{\partial \mathrm{B}}{\partial \mathrm{z}}\right)^{-2}$\cite{valahu2021robust}. Equations \ref{eq-gamma} and \ref{eq-min-sensitivity-theory} therefore show that the sensitivity in the current implementation of our electrometer becomes independent of both the secular frequency and the magnitude of the magnetic field gradient. However, this is not a universal scaling law, meaning modifications to the hardware of the sensor would improve the measured $\mathrm{S_{min}}$ and further increase the bandwidth of the sensor. These include reducing the PSD of voltage noise on the electrodes, replacing the existing low-pass filter (LPF) with one that has a larger roll-off rate and a lower cutoff frequency, or by using a voltage source that enables a different scaling of $\mathrm{T_2}$ with $\frac{\partial \mathrm{B}}{\partial \mathrm{z}}$ and $\nu_\mathrm{z}$. Additionally, the time penalty associated with phase matching electric field signals across the input capacitor leads to an increase in $\mathrm{t_m}$, increasing the minimum achievable sensitivity in the current experimental hardware (see methods). Using an in-vacuum antenna rather than a DC electrode as electric field source would avoid the need for capacitive coupling of electric field signals, thus leading to immediate improvements of $\mathrm{S_{min}}$.\par 
 Measured sensitivities can be further improved through hardware modifications of the quantum sensor. Extending the coherence time by reducing the voltage noise on the electrodes, in combination with dynamical decoupling techniques, would enable the use of first order magnetic field sensitive transitions as well as larger magnetic field gradients. Additionally, using a trapped ion with a large charge-to-mass ratio such as \ce{^{25}Mg^+} or \ce{^{9}Be^+} instead of \yb{} would further improve achievable sensitivities. For example, using the first order magnetic field sensitive $\ket{\mathrm{F=2,m_F=-2}}$ to $\ket{\mathrm{F=1,m_F=-1}}$ transition in the $\mathrm{S_{\frac{1}{2}}}$ hyperfine manifold of \ce{^{9}Be^+}, in a system with $\frac{\partial \mathrm{B}}{\partial \mathrm{z}}=\SI{200}{\tesla \meter^{-1}}$, would result in AC sensitivities of $<\SI{5e-9}{\volt \meter^{-1} \hertz^{-\frac{1}{2}}}$ for an evolution time of $\tau_\mathrm{opt}=\SI{170}{\milli \second}$ (and $\mathrm{T}_2=2\tau_\mathrm{opt}$). A further reduction in sensitivity by a factor of $\frac{1}{\sqrt{\mathrm{N}}}$ can also be achieved by increasing the number of quantum systems, N, within the sensor and evolving these in parallel. 
 \par 
Miniaturisation, portability, and hardware complexity are also important considerations for deploying quantum sensors in the field~\cite{schwindt2016highly}, and to ensure optimal positioning of the sensor relative to electric field sources. As the sensor presented in this work operates in-vacuum, sensor placement relative to a signal source may be more challenging for some applications. However, the development of compact ion-trapping systems is a well established area of research, with significant advancements being made in vacuum system miniaturisation~\cite{aikyo2020vacuum, Fernandez2023fibre}. Additionally, our scheme does not require cryogenic cooling of the hardware, which reduces portability constraints.\par 
In addition to improving sensitivities and portability, hardware modifications can broaden the range of applications of the sensor. A system that allows for independent tuning of the confinement strength along each axis of vibration can be used for the vector sensing of electric fields. Switchable static magnetic field gradients as described in \cite{siegele2022fabrication} could also be used to realise a hybrid magnetic field and electric field sensor, where this sensor has an identical measurement bandwidth for both magnetic and electric fields. Furthermore, our electric field sensor is compatible with entanglement-enhanced sensing techniques. Static magnetic field gradient entanglement schemes for trapped ions using long-wavelength radiation~\cite{mintert2001ion,weidt2016trapped} can be implemented, and could allow for the sensor to reach sensitivities below the standard quantum limit.\par 
%
%
\section*{Acknowledgements}
\label{sec-Acknowledgements}
This work was supported by the U.K. Engineering and Physical Sciences
Research Council via the EPSRC Hub in Quantum Computing and
Simulation (EP/T001062/1), the U.K. Quantum Technology hub for Networked Quantum Information Technologies (No. EP/M013243/1), the
European Commission’s Horizon-2020 Flagship on Quantum Technologies Project No. 820314 (MicroQC), the U.S. Army Research Office
under Contract No. W911NF-14-2-0106 and Contract No. W911NF-21-1-
0240, the Office of Naval Research under Agreement No. N62909-19-1-
2116 and the University of Sussex. F.B. and M.M. acknowledge the support from
the Engineering and Physical Sciences Research Council (EP/S021582/1)
via the Centre for Doctoral Training in Delivering Quantum Technologies
at the University College London.\par 
\section*{Author Contributions}
\label{sec-Authcont}
F.B. and C.K. contributed equally to this work, performed the experiments and analysed the data. C.H.V. conceived the idea. C.H.V. and M.M. wrote the experimental control software. S.W. and W.K.H. supervised this work. F.B., C.K. and C.H.V. wrote the manuscript. All authors discussed the results and contributed to the manuscript.\par 
\clearpage
\section*{Methods}
\subsection*{Transduction parameter}
\label{app:transduction_parameter}
We consider the dynamics of a string of $\mathrm{N}$ trapped ions perturbed by an external electric field, which results in a force $\delta \mathrm{F_j(t)} = -\mathrm{q} \delta \mathrm{E_j(t)}$ on ion $\mathrm{j}$. Restricting ourselves to a single direction without loss of generality, the Lagrangian of this system is~\cite{james1997quantum}
\begin{align}
    \mathrm{L} = & \frac{\mathrm{m}}{2} \left( \sum_{\mathrm{p}=1}^\mathrm{N} (\dot{\mathrm{Q}}_\mathrm{p(t)})^2 - \nu_\mathrm{p}^2 \mathrm{Q_p}^2(\mathrm{t})\right) + \nonumber \\
    & \mathrm{q} \mathrm{Q_p(t)} \sum_{\mathrm{j}=1}^\mathrm{N} \mathrm{b_j^{(p)}} \delta \mathrm{E_j(t)},
\end{align}
where $\nu_\mathrm{p}$ are the normal mode frequencies and $\mathrm{b_j^{(p)}}$ describes how strongly ion $\mathrm{j}$ couples to the mode $\mathrm{\mathrm{p}}$. The normal modes of motion, $\mathrm{Q_p(t)}$, are related to small displacements of the ion, $\delta \mathrm{r(t)}$ of Eq.\ref{eq:transduction_parameter}, via
\begin{equation}
    \mathrm{Q_p(t)} = \sum_{\mathrm{j}=1}^\mathrm{N} \mathrm{b_j}^{(\mathrm{p})} \delta \mathrm{r(t)}.
\end{equation}
The equation of motion of the $\mathrm{p^{th}}$ normal mode is found from the Lagrangian using the relation $\frac{\mathrm{d}}{\mathrm{dt}}(\frac{\partial \mathrm{L}}{\partial \dot{\mathrm{Q}}_\mathrm{p}}) = \frac{\partial \mathrm{L}}{\partial \mathrm{Q_p}}$, resulting in

\begin{equation} \label{eq:eq_motion}
    \ddot{\mathrm{Q}}_\mathrm{p}\mathrm{(t)} +  \nu_\mathrm{p}^2 \mathrm{Q_p(t)} = \frac{\mathrm{e}}{\mathrm{M}} \sum_{\mathrm{j}=1}^\mathrm{N} \mathrm{b_j^{(p)}} \delta \mathrm{E_j(t)}.
\end{equation}
Without loss of generality, we restrict ourselves to a single ion chain, $\mathrm{N}=1$, and consider the centre-of-mass motion along the z-axis. After setting $\mathrm{p}= \mathrm{z}$ and $\mathrm{b}_1^{(1)} = 1$, Eq.~\ref{eq:eq_motion} becomes
\begin{equation} \label{eq:eq_motion_simplified}
    \ddot{\mathrm{Q}}_\mathrm{z}(\mathrm{t}) +  \nu_\mathrm{z}^2 \mathrm{Q_z(t)} = \frac{\mathrm{e}}{\mathrm{M}}\delta \mathrm{E(t)}.
\end{equation}
This corresponds to the equation of a driven harmonic oscillator. Taking the Fourier transform, Eq.~\ref{eq:eq_motion_simplified} becomes

\begin{equation} \label{eq:fourier_transform_ion_motion}
    \hat{\mathrm{Q}}_\mathrm{p}(\omega) = \frac{\mathrm{e}}{\mathrm{m}(\nu_\mathrm{z}^2 - \omega^2)} \delta\hat{\mathrm{E}}(\omega),
\end{equation}
where $\hat{\cdot}$ denotes the Fourier transform. For $\mathrm{N}=1$ ion, we find $\mathrm{Q_p(t) = \delta r(t)}$ and Eq.~\ref{eq:fourier_transform_ion_motion} becomes
\begin{equation}
\label{eq:fourier_transform_ion_motion_deltar}
    \delta\hat{\mathrm{r}}(\omega) = \frac{\mathrm{e}}{\mathrm{m}(\nu_\mathrm{z}^2 - \omega^2)} \delta\hat{\mathrm{E}}(\omega).
\end{equation}

In the limit $\nu_\mathrm{z} \gg \omega$, Eq.~\ref{eq:fourier_transform_ion_motion_deltar} reduces to 
\begin{equation}
    \delta\hat{\mathrm{r}}(\omega) = \frac{\mathrm{e}}{\mathrm{m}\nu_\mathrm{z}^2} \delta\hat{\mathrm{E}}(\omega),
\end{equation}
from which one can retrieve the expression of Eq.~\ref{eq:transduction_parameter}.\par
\subsection*{Experimental setup}
\begin{figure*}[ht]
    \centering
    \includegraphics[width=0.9\textwidth]{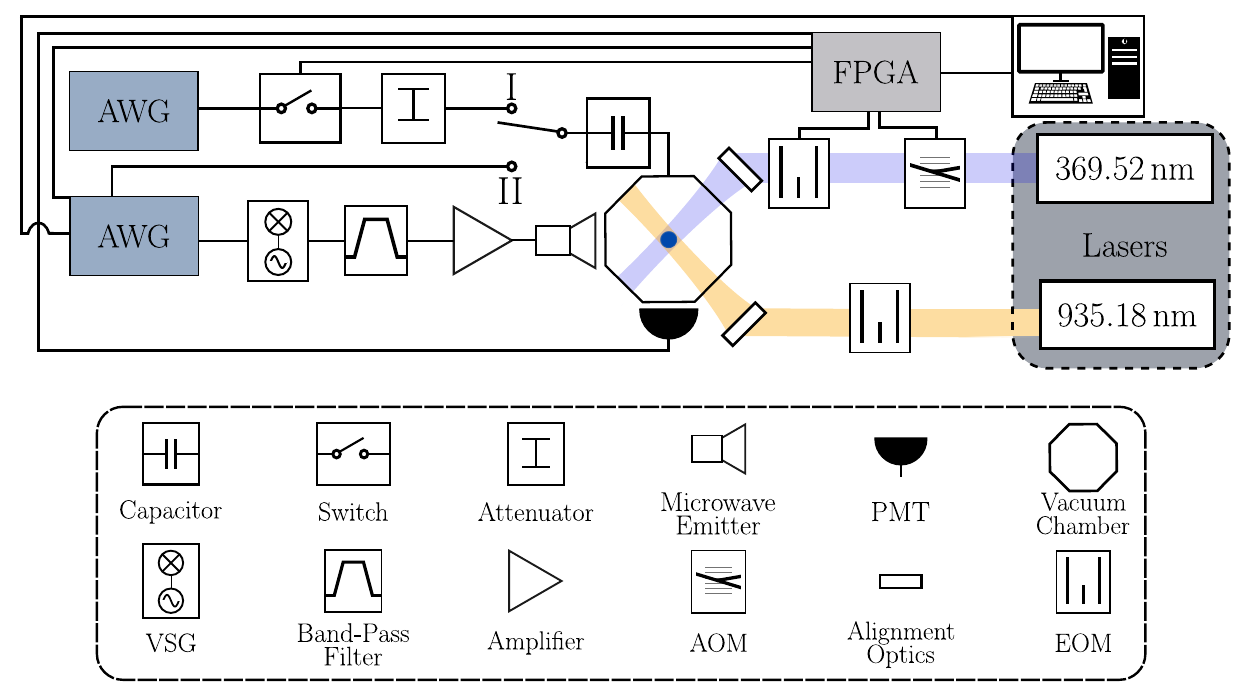}
    \caption{\textbf{Experimental Setup.} Electric field sensing configurations, coherent control, optical and electric field signal chains for the operation of the quantum sensor. Coherent control is achieved using triggered microwave pulses generated by amplitude modulation of an RF signal from a two-channel AWG with a microwave carrier using a VSG. The microwave tone is amplified and emitted into the vacuum chamber using a microwave horn. The second channel of this AWG provides the electric field signals for AC and DC sensing. These signals are synchronously coupled into the quantum sensor in configuration II. Configuration I shows the setup for rotating frame relaxometry. Here, a signal is continuously output using a second AWG. Interaction with the spin state is toggled using an RF switch. The signal is then attenuated and capacitively coupled onto the electrode. Doppler cooling, optical pumping and state detection of the ion are achieved by modulating a \SI{369.52}{\nano \meter} laser beam using an acousto-optic modulator (AOM) and an electro-optic modulator (EOM). An EOM in the \SI{935.18}{\nano \meter} beam allows for efficient repumping. The photo-multiplier tube (PMT) is used for fluorescence detection of the spin state.}
    \label{fig-expsetup}
\end{figure*}

Figure~\ref{fig-expsetup} shows a schematic of the experimental setup used in this work. The ion trap is mounted inside a vacuum chamber maintained at an average pressure of $\SI{2.4e-11}{\milli \bar}$. The ion is Doppler cooled using a $\SI{369.52}{\nano \meter}$ laser that is red-detuned from the $\mathrm{^2S_{\frac{1}{2}} \ket{F=1}}$ to $\mathrm{^2P_{\frac{1}{2}} \ket{F=0}}$ transition. The laser beam is double-passed through an acousto-optic modulator (AOM) to allow for fine frequency and amplitude control via FPGA. An electro-acoustic modulator (EOM) is also utilised to generate $\SI{2.11}{GHz}$ sidebands for state preparation. These sidebands allow the population to be driven into the $\mathrm{^2P_{\frac{1}{2}} \ket{F=1}}$ state via optical pumping, after which it decays into the $\ket{\downarrow} = \mathrm{^2S_{\frac{1}{2}} \ket{F=0}}$ ground state. Population that is off-resonantly driven into the $\mathrm{^2S_{\frac{1}{2}} \ket{F=0}}$ during Doppler cooling is returned to the cooling cycle by continuously applied microwaves near $\SI{12.64}{GHz}$. Population can also exit the Doppler cooling cycle by decaying into the $^2\mathrm{D}_{\frac{3}{2}}$ manifold, where a $\SI{935.18}{\nano \meter}$ re-pump laser applied on the $^2\mathrm{D}_{\frac{3}{2}}$ to $^3\mathrm{D[3/2]}_{\frac{1}{2}}$ transition returns population into the $\mathrm{^2S_\frac{1}{2}\ket{F=1}}$ state. The re-pump laser is also modulated by an EOM at $\SI{3.07}{GHz}$ to improve re-pumping efficiencies. The microwaves are generated by a Keysight E8267D PSG Vector signal generator (VSG), which produces a carrier signal of $\SI{12.54}{GHz}$. This is then mixed with radio frequency (RF) pulses near $\SI{100}{\mega \hertz}$ generated via a Keysight M8190A two-channel arbitrary waveform generator (AWG), which is then amplified and emitted via an external microwave emitter to allow for coherent manipulation of the spin state. Measurement of the spin is carried out using a state dependent fluorescence scheme as described in \cite{olmschenk2007manipulation}, where the average SPAM error is $\eta=1.8\times10^{-2}$. The voltage signals used for measurements of the AC and DC sensitivity are directly applied into the capacitor from the second channel of the M8190A AWG. For electric field noise measurements the white-noise waveform is generated with an Agilent 33522A AWG. The white noise signal is attenuated by two \SI{30}{\decibel} RF attenuators and its output is controlled with an external RF switch.\par 
\subsection*{Gradient measurement}
\begin{figure}[h!]
    \centering
    \includegraphics[width=\columnwidth]{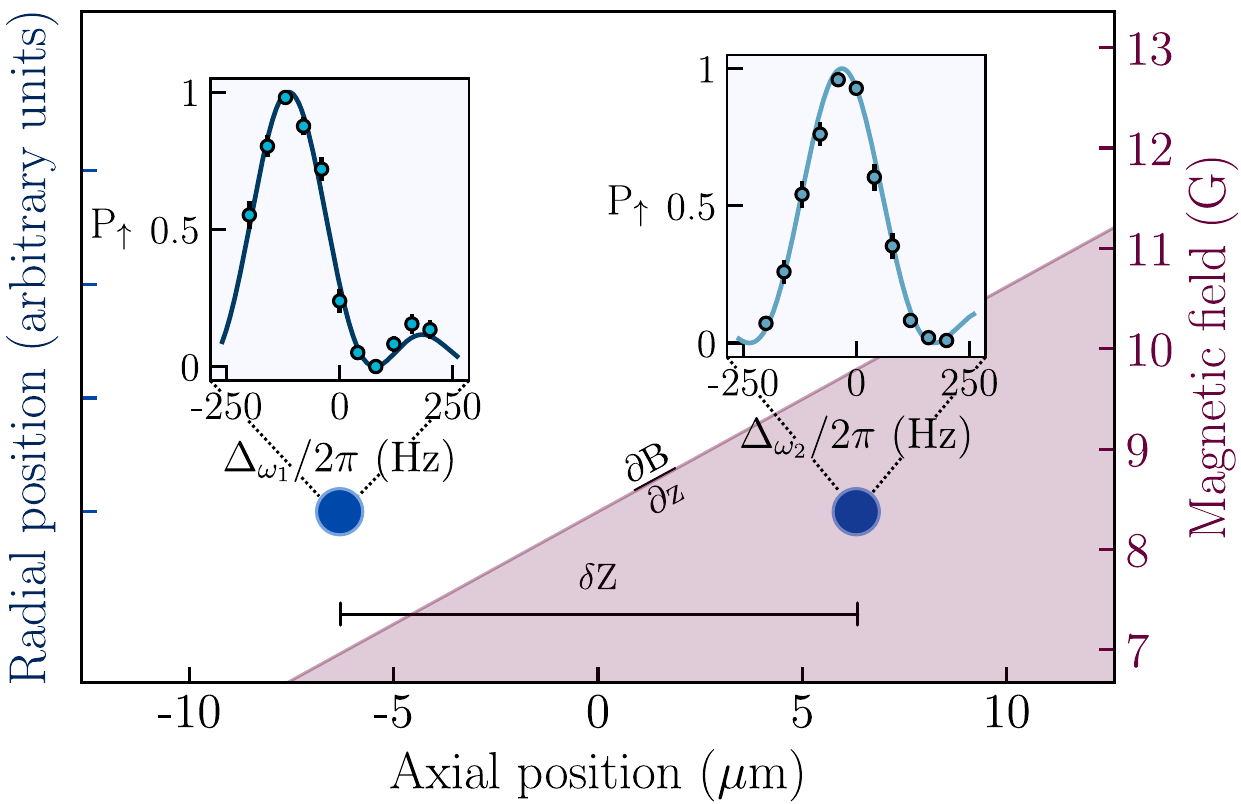}
    \caption{\textbf{Measurement of the magnetic field gradient.} The blue markers represent two co-trapped \yb{} ions that form an ion crystal in the $\mathrm{y}-\mathrm{z}$-plane. The left y-axis represents the radial ($\mathrm{y}$) position of the ions. The axial ($z$) separation of the ions is \SI{12.64(1)}{\micro \meter}, and is symmetric about the single ion equilibrium location in the axial direction of the trap. The increasing value of absolute magnetic field with axial position is represented by the threshold between the white and purple shaded regions. Values of magnetic field for a given axial position are displayed on the right y-axis. The insets show measurements and least-squares fits of the spin state transition frequency of each ion. The horizontal axis of the insets indicate the detuning of the applied microwave pulse, relative to probing frequencies near \SI{12.64}{\giga \hertz} corresponding to the spin transition of ion 1, $\omega_1$ (left inset), and ion 2, $\omega_2$ (right inset).}
    \label{fig-gradientmeas}
\end{figure}

The magnetic field gradient strength along the axial direction can be calculated by measuring the transition frequencies of two co-trapped \yb{} ions. As the splitting of the \yb{} spin states is dependent on the strength of the magnetic field at the position of the ion, the magnetic field gradient in the axial direction is given by %
\begin{equation}
\label{eq-gradientcalc}
\frac{\partial \mathrm{B}}{\partial \mathrm{z}} = \frac{\mathrm{B_{2} - B_{1}}}{\delta \mathrm{Z}},
\end{equation}
where $\mathrm{B_1}$ and $\mathrm{B_{2}}$ are the magnetic field strengths at the location of each ion, and $\delta \mathrm{Z}$ is the ion separation (see figure~\ref{fig-gradientmeas}). The ion separation is a result of the mutual coulomb repulsion between the ions and the oppositely acting axial confinement force. $\delta \mathrm{Z}$ given by~\cite{james1997quantum},
\begin{equation}
\label{eq:separation}
\delta \mathrm{Z}=\left(\frac{\mathrm{e}^{2}}{4\pi\epsilon_{0}\mathrm{m}\nu^2_z}\right)^2\frac{2.018}{\mathrm{N}^{0.559}},
\end{equation}
where $\nu_\mathrm{z}$ is the axial vibrational centre of mass (COM) frequency, m is the mass of a single charged particle, and N is the number of ions in the crystal. We measure $\nu_\mathrm{z}/2\pi=\SI{161.191(8)}{\kilo \hertz}$ via the ``tickling'' method~\minus~an AC electric field is applied to the trap using an external RF coil, which excites the axial motion of the ion crystal when the applied frequency is resonant with the axial vibrational frequency, leading to a measurable decrease in ion fluorescence due to the Doppler shift. We then compute $\mathrm{\delta Z = \SI{12.64(1)}{\micro \meter}}$ from equation~\ref{eq:separation}. \par
The magnetic field at each ion is calculated by measuring the magnetic field dependent transition frequency of each ion, shown in the inset plots of figure~\ref{fig-gradientmeas}. From these measurements we find $\mathrm{B}_1 = \SI{7.1328(8)}{G}$ and $\mathrm{B}_2 = \SI{9.9655(5)}{G}$. Finally, from equation~\ref{eq-gradientcalc}, the magnetic field gradient strength is $\frac{\partial \mathrm{B}}{\partial \mathrm{z}}=\SI{22.41(1)}{\tesla \per \meter}$.\par
\subsection*{Calibrating $\alpha$ and $\gamma$}
\begin{figure}[h]
    \centering
    \includegraphics[width=\columnwidth]{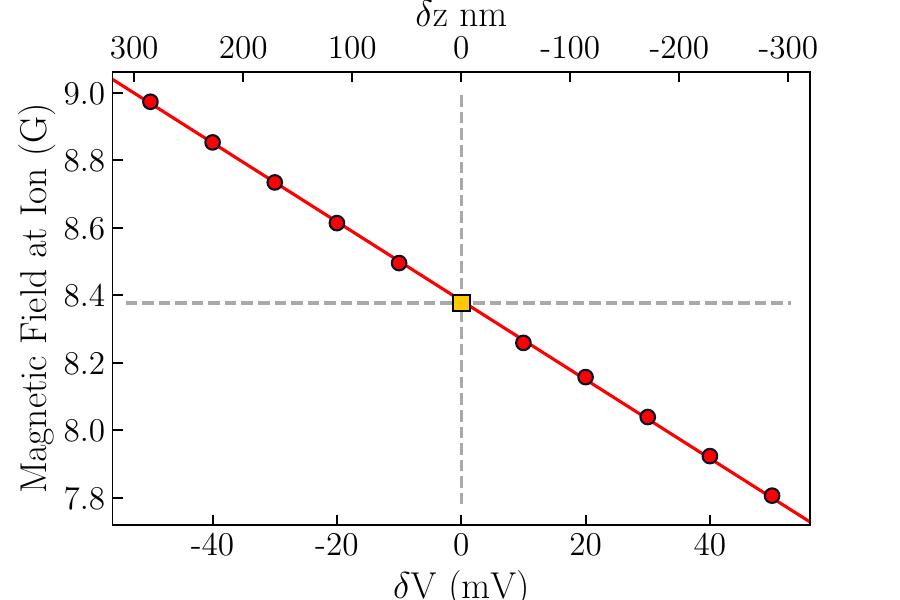}
    \caption{\textbf{Calibrating the geometric factor.} 
    Measurement of the shift in resonance frequency of the two-level system after applying a static voltage offset $\mathrm{\delta V}$ to the E1 electrode. The red circles indicate the magnetic field determined from spin state transition frequency measurements at different values of $\mathrm{\delta V}$, whilst the yellow square represents the magnetic field corresponding to a measurement of the unperturbed ($\delta \mathrm{V}=0$) transition frequency. The data are fitted to a straight line using a least-squares fit, shown in red. The top axis shows the axial displacement of the ion for a given $\mathrm{\delta V}$, which is calculated using the previously measured value of $\alpha$.}
    \label{fig-a_kmeas}
\end{figure}

The geometric factor of an electrode, $\alpha$, relates the electric field at the position of the ion to the voltage applied to the electrode, and is defined as
\begin{equation} \label{eq-geometric-factor}
\alpha = \frac{\partial \mathrm{E}}{\partial \mathrm{V}} = \frac{\partial \omega}{\partial \mathrm{V}} \frac{\partial \mathrm{E}}{\partial \mathrm{z}}\left(\frac{\partial \mathrm{B}}{\partial \mathrm{z}}\frac{\partial \omega}{\partial \mathrm{B}}\right)^{-1},
\end{equation}
where $\frac{\partial \mathrm{E}}{\partial \mathrm{z}}=\frac{\mathrm{m} \nu^2_{\mathrm{z}}}{\mathrm{e}}$. We calibrate $\alpha$ by first measuring the change in magnetic field at the ion due to a change in the voltage applied to the E1 electrode ($\frac{\partial \mathrm{B}}{\partial \mathrm{V}}$) using the second order sensitive spin state transition frequency 
 and $\nu_z/2\pi=\SI{161.191(8)}{\kilo \hertz}$ (see figure~\ref{fig-a_kmeas}). The measurement is performed with a single \yb{} ion by applying a voltage $\mathrm{V_0 + \delta V}$ to the electrode, where $\mathrm{V_0 = \SI{1.75}{\volt}}$ is the static voltage contributing to the axial confining potential and $\delta \mathrm{V}$ is an offset that is varied from $-\SI{50}{\milli \volt}$ to $+\SI{50}{\milli \volt}$. We extract the value of $\frac{\partial \mathrm{B}}{\partial \mathrm{V}}$ from a least squares fit to a straight line of the magnetic field measurements for each voltage offset. From this, we then determine $\frac{\partial \omega}{\partial \mathrm{V}}= \frac{\partial \mathrm{B}}{\partial \mathrm{V}} \frac{\partial \omega}{\partial \mathrm{B}} = -\SI{382e3}{\radian \per \volt}$. The geometric factor is finally calculated from equation~\ref{eq-geometric-factor}, giving $\alpha=\SI{-95.64(4)}{\per \meter}$.\par 
The transduction parameter is found using $\gamma = \frac{1}{\alpha} \frac{\partial \omega}{\partial \mathrm{V}}=\left(\frac{\partial \mathrm{V}}{\partial \mathrm{E}}\frac{\partial \omega}{\partial \mathrm{V}}\right)$. For the second order magnetic field sensitive transition we measure $\gamma=\SI{3998(2)}{\radian \meter \per \volt}$.\par 
\subsection*{Electric field sensing protocol}
\begin{figure}[h]
    \centering
    \includegraphics[width=\columnwidth]{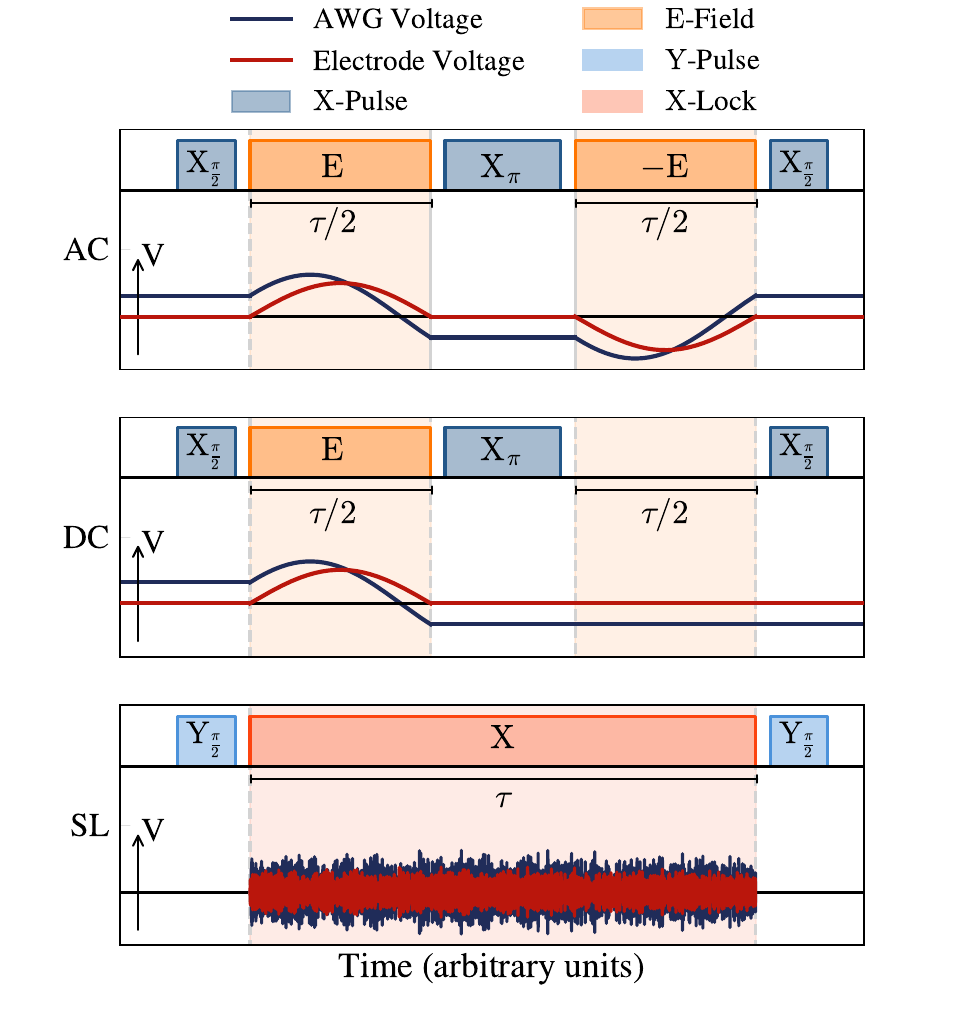}
    \caption{\textbf{AC, DC and spin-locking (SL) pulse sequence diagram and time evolution of input signal.} Blue lines represent the AWG voltage output, whilst red lines show the voltage evolution on the in-vacuum electrode. Note that the voltage on the electrode is attenuated and phase-offset relative to the AWG voltage. The AC sensing technique is characterised by applying a full oscillation of an AC signal at frequency $\frac{1}{\tau}$ onto the electrode. Sensitivity to DC signals is characterised by applying a half-oscillation of an AC signal at frequency $\frac{1}{\tau}$ onto the E1 electrode of the ion trap. For spin-locking, noise resonant with the spin-locking Rabi frequency is applied onto the electrode. The input signal exhibits a frequency dependent phase shift and a frequency dependent attenuation across the input capacitor. For AC and DC sensing the offsets are pre-compensated, as can be seen by the dark-blue line. For spin-locking a continuous signal is switched into, thus pre-compensation is not applied.}
    \label{fig-ACDCPSonly}
\end{figure}

For the sensing of AC fields, we follow the pulse sequence protocol outlined in~\cite{degen_2017} and illustrated in figure~\ref{fig-ACDCPSonly}. The AC sensing sequence is realised by first initialising the two-level system into the $\ket{+} = \frac{1}{\sqrt{2}}(\ket{\downarrow}+\ket{\uparrow})$ state using a $\frac{\pi}{2}$-pulse. The superposition state then evolves under an electric field perturbation for a time $\frac{\tau}{2}$. A $\pi$-pulse reorients the spin along the equator of the Bloch sphere, before the quantum state again evolves under the electric field perturbation for a time $\frac{\tau}{2}$. A final $\frac{\pi}{2}$-pulse maps the state population into the $\sigma_\mathrm{z}$-basis for detection. Using this pulse-sequence, the sensitivity of the spin state transition frequency is maximised for AC signals oscillating at a frequency of $\tau^{-1}$.\par
The DC sensing experiments also use an echo-type pulse sequence, whose benefits are twofold. Firstly, the coherence time of the sensor is greatly extended when compared to that of the Ramsey type sequence, which allows for increased sensitivities. Secondly, the refocusing $\pi$-pulse also compensates for detuning errors in the microwave pulses. The pulse sequence is also illustrated in figure~\ref{fig-ACDCPSonly}. The pulse sequence begins with a $\frac{\pi}{2}$-pulse to initialise the spin into the $\ket{+} = \frac{1}{\sqrt{2}}(\ket{\downarrow}+\ket{\uparrow})$ state. DC signals cannot be applied through a capacitor. The low-pass filter (LPF) signal chain of the DC electrode is also not suitable for a fast application of DC square pulses during the sensing pulse sequence, as the LPF would significantly attenuate and distort the signal. Therefore, in order to quantify the sensor's response to DC signals, we apply an AC signal of frequency $\tau^{-1}$ for the duration of the first $\frac{\tau}{2}$ delay time. This corresponds to an equivalent DC voltage on the electrode of $\mathrm{V_{DC} = \frac{2}{\pi} V_{PK}}$ where $\mathrm{V_{PK}}$ is the amplitude of the applied signal. $\frac{2}{\pi} \mathrm{V_{PK}}$ is the average voltage over the half-oscillation of the AC waveform. The applied time-varying pulse therefore causes the spin state to accumulate the same amount of phase, $\phi$, as a square DC pulse of amplitude $\mathrm{\frac{2}{\pi}V_{PK}}$ applied for a duration $\frac{\tau}{2}$ based on the equation relating phase accumulation to the detuning of the spin transition: $\phi=\int_0^\frac{\tau}{2}  \gamma \alpha \delta \mathrm{V}\mathrm{(t) dt}$. The refocusing $\pi$-pulse is then applied, followed by the second $\frac{\tau}{2}$ delay time, during which no additional voltage signals are applied to the electrode, followed by a final $\frac{\pi}{2}$-pulse.\par 
In addition to the electric field interaction time $\tau$ the second relevant time parameter from equation \ref{eq-min-sensitivity-theory} is $\mathrm{t_m}$ which breaks down as follows for our experimental implementation: DC offset application delay time: \SI{50}{\milli \second} (see below section), Doppler cooling and detection: $\SI{14.599}{\milli \second}$, state preparation and microwave pulses: \SI{2.155}{\milli \second}, data processing and FPGA delays: \SI{85}{\micro \second}, total: $\mathrm{t_m}=\SI{66.839}{\milli \second}$.\par 
\subsection*{Capacitive coupling of AC signals}
Due to the absence of an in-vacuum antenna, the electric field signals measured by the trapped ion are emitted from an in-vacuum end-cap electrode, which also generates a DC confinement electric field. Voltage waveforms are generated using an AWG, and capacitively coupled onto the electrode across a $\SI{220}{pF}$ capacitor. Due to their frequency dependent impedance, capacitors act as high-pass filters, thereby attenuating lower frequency signals more strongly. The fixed response time of a capacitor will also shift the phase of AC signals that are applied across it. The shift in phase of the AC signal can, if unaccounted for, affect the total coherent phase $\phi$ that is accumulated by the spin states. In order to achieve an optimal measurement of the sensitivity of our experimental system, it is necessary for the electric field signal at the ion to be in-phase with the Hahn-echo sensing pulse sequence. This is because $\phi$ is the difference between the coherent-phase accrued in the first and second interaction time $\frac{\tau}{2}$. An electric field signal that is not in phase with the Hahn-echo sequence will therefore lead to a reduction in measured sensitivity. References~\cite{michl2019robust} and~\cite{degen_2017} provide further information on this effect.\par
We measure the phase shift on signals applied across the capacitor for the span of frequencies used in the AC and DC sensing experiments using an oscilloscope. Based on these measurements, we then pre-compensate the signal that is applied across the capacitor by applying an inverse phase shift, negating the effect of the capacitor on the phase of the voltage waveform. This ensures that the voltage on the electrode, and therefore the electric field signal at the ion, is in phase with the Hahn-echo sequence.\par
Shifting the phase of the voltage waveform introduces a discontinuity into the signal. This manifests as a sudden change in the voltage across the capacitor from $0$ to $\mathrm{V_{\Phi} = V_A\sin{\Phi}}$ where $\Phi$ is the phase of the AC voltage signal. Given that the current across a capacitor is defined as $\mathrm{I = C\frac{dV}{dt}}$, where $\mathrm{C}$ is the capacitance of the capacitor, the high rate of change of voltage induces a large current flow across the capacitor, which introduces additional coherent phase offsets of the superposition state. To suppress this unwanted perturbation, we apply a DC voltage offset of $\mathrm{V}_{\Phi}$ into the capacitor in the time prior to the initialisation of the $\ket{+}$ state, which minimises the sudden voltage spike across the capacitor from the phase-shifted AC voltage waveform. To ensure that the sensor reaches steady state prior to the application of the AC electric field signal, an additional \SI{50}{\milli \second} delay is added between the application time of the DC offset and the first resonant microwave pulse. This makes up the majority of the $\mathrm{t_m}$ time, which is broken down in the previous section. The pre-compensation technique for the AC and DC sensing pulse sequences is visualised in figure~\ref{fig-ACDCPSonly}, which illustrates both the AWG and in-vacuum electrode voltage evolution throughout the experimental pulse-sequence.\par 
We also measure the frequency dependent attenuation of the capacitor using an oscilloscope. We determine the transfer function of the capacitor by fitting a Butterworth high-pass filter to these data. We then determine the total attenuation of the electric field signal for a given frequency $\tau^{-1}$.\par 

\subsection*{Determination of the coherence time}
\begin{figure}[h!]
    \centering
    \includegraphics[width=\columnwidth]{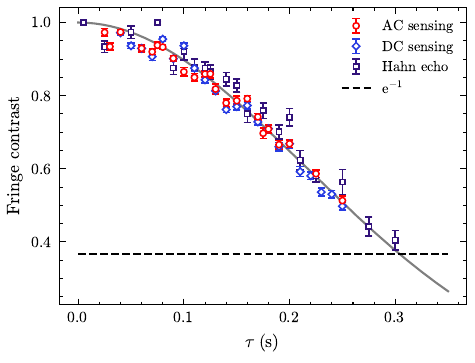}
    \caption{\textbf{Coherence time measurement of the second order sensitive clock states.} The spin states used for the experimental demonstration of AC and DC sensing are defined by the $\mathrm{\ket{\downarrow} = \ket{F = 0, m_F = 0}}$ and $\mathrm{\ket{\uparrow} = \ket{F = 1, m_F = 0}}$ energy levels. The fringe contrasts associated with each of the AC sensing, DC sensing, and Hahn-echo experiments are shown for a range of free evolution times $\tau$. The black dashed line indicates the $1/\mathrm{e}$ threshold. The grey line is a least squares fit of these measurements to a Gaussian decay function, corresponding to a coherence time of $\mathrm{T}_2 = \SI{304(3)}{\milli \second}$.}
    \label{fig-t2coherence}
\end{figure}

We measure the coherence time of the two-level system using a Hahn-echo experiment. The spin is initialised in the $\ket{\downarrow}$ state, after which a $\frac{\pi}{2}$-pulse rotates the spin into the $\ket{+\mathrm{X}}$ eigenstate. A refocusing $\pi$-pulse is applied in-between two free evolution periods of duration $\frac{\tau}{2}$. A final $\frac{\pi}{2}$-pulse maps the state into the $\sigma_\mathrm{z}$-basis for detection. Varying the phase of the final pulse from $-2\pi$ to $2\pi$ results in sinusoidal fringes in the probability of measuring $\ket{\uparrow}$. As the free evolution time is increased, decoherence leads to a reduction in the amplitude of these fringes, where the coherence time $\mathrm{T}_2$ is then given by the point at which the fringe contrast reaches $\mathrm{e}^{-1}$. As the AC and DC sensing experiments are also based on the Hahn-echo sequence, the fringe amplitudes from these experiments can also be utilised for the coherence time measurement. The fringe amplitudes of these three experiments against the free evolution time are shown in figure~\ref{fig-t2coherence}. These data are then aggregated and fitted to a Gaussian decay function of the form $\chi^{-1}(\mathrm{t}) = \exp(-\frac{\mathrm{t}^2}{\mathrm{T}_{2}^2})$ using a least squares fit, yielding a coherence time of $\mathrm{T}_2=\SI{304(3)}{\milli \second}$.

%
%

\clearpage
\bibliography{biblio}
\bibliographystyle{unsrt}

\end{document}